\documentclass[a4paper]{jpconf}

\usepackage[british]{babel}
\usepackage{amsmath,amssymb,bbm}
\usepackage{multicol}

\newcommand{\nocomma}{}

\newcommand{\tmop}[1]{\ensuremath{\operatorname{#1}}}
\newcommand{\tmrsup}[1]{\textsuperscript{#1}}

\newcommand{\um}{{-}}
\newcommand{\upl}{{+}}
\begin{document}

\bibliographystyle{unsrt}

\makeatletter
\@definecounter{equation}
\def\equation{$$\refstepcounter{equation}}
\def\endequation{\eqno \hbox{\@eqnnum}$$\@ignoretrue}
\def\@eqnnum{{\normalsize\normalfont \normalcolor (\theequation)}}
\makeatother

\title{
  The generalized $t$-$V$ model in one dimension
}

\author{Marcin Szyniszewski\tmrsup{1,2}, Evgeni Burovski\tmrsup{1}}

\address{\tmrsup{1} Physics Department, Lancaster University, Lancaster, LA1~4YB, UK}

\address{\tmrsup{2} NoWNano DTC, University of Manchester, Manchester, M13~9PL, UK}

\ead{mszynisz@gmail.com, e.burovski@lancaster.ac.uk}

\begin{abstract}
	We develop a systematic strong coupling approach for studying an extended
	$t$-$V$ model with interactions of a finite range. Our technique is not
	based on the Bethe ansatz and is applicable to both integrable and non-integrable models. 
	We illustrate our technique by presenting analytic results for the ground state
	energy (up to order 7 in $t/V$), the current density and density-density correlations 
	for integrable and non-integrable models with commensurate filling factors.
	We further present preliminary numerical results for incommensurate non-integrable
	models.
\end{abstract}

\section{Introduction}

Low-dimensional materials are of very high interest at present due to
their exceptional electronic properties. Furthermore, the effects of
interactions are enhanced in low dimensions, which leads to a variety
of highly non-trivial quantum phases. In one spatial dimension, prime
examples are the Luttinger liquid and the Mott insulating phase
{\cite{Giamarchi}}.
While low-energy properties of these phases are well described by
effective field theory techniques, calculating the parameters of the
effective theories from first principles is typically challenging. One
well established and often used method is based on the Bethe ansatz,
which, in principle, can provide an exact solution for integrable systems.

The main limitation of the Bethe ansatz is its rather limited applicability:
it can only be applied to models of very high symmetry (as integrability
implies an infinite number of conserved quantities). Generalizing
Bethe-ansatz-based calculations to non-integrable systems does not seem possible.

In this work, we study a specific lattice model, the so-called $t$-$V$ model,
which has both integrable and non-integrable regimes. A rather elegant way
of investigating the infinite coupling limit of this model was developed
in Ref. {\cite{Gomez-Santos1993}} and a family of Mott insulating phases was found.
We use a variant of the strong coupling expansion {\cite{Hamer1979}}, mainly used for
the investigation of lattice field theories {\cite{Crewther1980,Hamer1997,Cichy2013}},
to extend and generalize the results of Ref. {\cite{Gomez-Santos1993}} to
large but finite couplings. Our method is insensitive to integrability (or
the lack of it) and we obtain ground state properties of the model as a series
in $t/V$ with minimal effort.

The rest of the paper is organized as follows. Firstly, we will present the model
in question. Then, we will present the strong coupling expansion (SCE) as a general
method for any quantum model. We will also investigate the generalized $t$-$V$ model
for Mott insulator densities using SCE and compare the results to the
previous work. Finally, we will conclude by sketching the future research into
this subject.

\section{Generalized $t$-$V$ model}

The Hamiltonian of long-range $t$-$V$ model of fermions on a one-dimensional
ring of size $L$ is as follows {\cite{Gomez-Santos1993}}:
\begin{equation}
  \hat{H} = - t \sum_{i = 1}^L \left( \hat{c}^{\dag}_i \hat{c}_{i + 1} +
  \text{h.c.} \right) + \sum_{i = 1}^L \sum_{m = 1}^p U_m \hat{n}_i \hat{n}_{i
  + m} \label{ham}
\end{equation}
where $\hat{c}_i$ is a fermionic operator on site $i$, $\hat{n}_i = \hat{c}_i^{\dag} \hat{c}_i$ 
is a particle number operator, $p$ is the maximum range of interactions and
$t$ and $U_m$ are kinetic and potential energies respectively.
We use periodic boundary conditions. For the range $p =
1$, the model is integrable and equivalent to the XXZ Heisenberg model after a
Jordan-Wigner transformation {\cite{Jordan1928}}. The model is non-integrable
for $p > 1$. We assume that $U_m \leqslant \frac{U_{m - 1} + U_{m + 1}}{2}$,
so that we are in the correct phase of the system, as described in Refs.
{\cite{Gomez-Santos1993,Schmitteckert2004,Mishra2011}}. Another assumption is
that the kinetic energy term will always be very small, $t \ll U_m$, and thus
we can rewrite the Hamiltonian as:
\begin{equation}
  \hat{H} = \hat{H}_0 + \lambda \hat{V}
\end{equation}
with $\hat{H}_0$ being the unperturbed Hamiltonian containing the potential
energy, $\hat{V}$ -- perturbation (kinetic energy) and $\lambda = t$ is a
small parameter.

\section{Strong coupling expansion}

In the article by Hamer {\cite{Hamer1979}}, he introduced a method to truncate
the basis according to how states were connected to the unperturbed initial
subspace. The method is to reorder the basis (usually this is the computer
basis), firstly writing the desired subspace of unperturbed states that we
want to approximate (0th step), then states connected to them (1st step), then
states connected to the 1st step states (2nd step) and so on. It is easy to
see that this results in a tri-block-diagonal Hamiltonian. We truncate the
basis to the step of our choice, resulting in smaller, truncated Hamiltonian,
which will describe the full system up to a specific perturbation order.
However, the truncated basis is still usually quite big, thus we will use an
altered version of this method, commonly used in the investigations of the
one-dimensional analogue of quantum electrodynamics -- the Schwinger model
{\cite{Crewther1980,Hamer1997,Cichy2013}}. 

The method is as follows. Firstly, let us select the desired initial subspace
of unperturbed states that we want to approximate. Usually that will be the
ground state, but if one is interested in the temperature dependance, that
could be first excited states, second excited states, etc. We will designate
states in this subspace by $| 0^i \rangle$, which means that we will treat it
as a 0th step of our SCE.

Secondly, to create states of the next step in SCE, we will act with
perturbation operator $\hat{V}$ on the states from previous step, $\hat{V} |
n^i \rangle$. $\hat{V} | n^i \rangle$ will be, in general, a linear
combination of states from orders $n \um 1$, $n$ and $n \upl 1$. It will not
include lower orders, because $\hat{V} | n^i \rangle$ is defined to not
include orders higher than $n \upl 1$, which means $\forall_{m > n \upl 1}
\langle m^j | \hat{V} | n^i \rangle = 0$ and $\forall_{n <
m \um 1} \langle n^i | \hat{V} | m^j \rangle = 0$. This shows that the
Hamiltonian in such a basis is tri-block-diagonal, as in the original Hamer
method. To properly define states in order $n \upl 1$, we have to separate
states in $\hat{V} | n^i \rangle$ according to their unperturbed energy -- the
states must be eigenstates of $\hat{H}_0$. Thus, in the end:
\begin{equation}
  \hat{V} | n^i \rangle = \sum_j C_j | n \um 1^j \rangle + \sum_k C_k | n^k
  \rangle + \sum_l | \widetilde{n \upl 1}^l \rangle \label{separation}
\end{equation}
where $C_j, C_k$ are normalization constants. The new states $| \widetilde{n
\upl 1}^l \rangle$ are not yet orthonormal to each other and to the previous
states. After Gramm-Schmidt orthonormalization they become:
\begin{equation}
  | n^j \rangle = C_{\tilde{n}, j} | \tilde{n}^j \rangle - \sum_{m = 1}^{n -
  1} \sum_{k = 1}^{k_{\max} (m)} C_{\tilde{n}, j ; m, k} | m^k \rangle -
  \sum_{k = 1}^{j - 1} C_{\tilde{n}, j ; n, k} | n^k \rangle
  \label{orthonormalization}
\end{equation}
where coefficient $C_{\tilde{n}, j}$ is normalization and other coefficients
include normalization and projection: $C_{\tilde{n}, j ; m, k} = C_{\tilde{n},
j} \langle m^k | \tilde{n}^j \rangle$.

If we continue this procedure infinitely long, we will not necessarily produce the
full basis. Thus, there may be states that are not producible by this procedure, which
we will call $| \alpha \rangle$, and which will form, together with states $|
n^i \rangle$, an orthonormal non-truncated basis of the system. However, we can
easily see that using (\ref{separation}) and then 
(\ref{orthonormalization}):
\begin{eqnarray}
  \langle \alpha | \hat{V} | n^i \rangle & = & \langle \alpha | \left( \sum_j
  C_j | n \um 1^j \rangle + \sum_k C_k | n^k \rangle + \sum_l | \widetilde{n
  \upl 1}^l \rangle \right) \\
%  & = & \langle \alpha | \sum_j C_j | n \um 1^j \rangle + \langle \alpha |
%  \sum_k C_k | n^k \rangle + \langle \alpha | \sum_l \frac{1}{C_{\widetilde{n
%  \upl 1}, l}} \times \nonumber\\
%  & \times & \left( | n \upl 1^l \rangle + \sum_{r = 1}^n \sum_{k =
%  1}^{k_{\max} (m)} C_{\widetilde{n + 1}, l ; r, k} | r^k \rangle + \sum_{k =
%  1}^{l - 1} C_{\widetilde{n \upl 1}, l ; n \upl 1, k} | n \upl 1^k \rangle
%  \right) \nonumber\\
  & = & \sum_j C_j \langle \alpha | n \um 1^j \rangle + \sum_k C_k \langle
  \alpha | n^k \rangle + \sum_l \frac{1}{C_{\widetilde{n \upl 1}, l}} \times
  \nonumber\\
  & \times & \left( \langle \alpha | n \upl 1^l \rangle + \sum_{r = 1}^n
  \sum_{k = 1}^{k_{\max} (m)} C_{\widetilde{n \upl 1}, l ; r, k} \langle
  \alpha | r^k \rangle + \sum_{k = 1}^{l - 1} C_{\widetilde{n \upl 1}, l ; n
  \upl 1, k} \langle \alpha | n \upl 1^k \rangle \right) \nonumber\\
  & = & 0 \nonumber
\end{eqnarray}
This proves that states $| \alpha \rangle$ are in fact part of a completely
different subspace of the Hamiltonian than states $| n^i \rangle$. Therefore,
eigenvalues of the desired subspace that we will be approximating will not
depend on $| \alpha \rangle$ and neither will any averages over states from
this subspace.

The Hamiltonian is now in the tri-block-diagonal form:
\begin{equation}
  \hat{H} = \left(\begin{array}{ccccc|c}
    \hat{E}_0 + \lambda \hat{V}_{00} & \lambda \hat{V}_{01} & 0 & 0 & \cdots &
    \\
    \lambda \hat{V}_{01}^T & \hat{E}_1 + \lambda \hat{V}_{11} & \lambda
    \hat{V}_{12} & 0 & \cdots & \\
    0 & \lambda \hat{V}_{12}^T & \hat{E}_2 + \lambda \hat{V}_{22} & \lambda
    \hat{V}_{23} &  & 0\\
    0 & 0 & \lambda \hat{V}_{23}^T & \hat{E}_3 + \lambda \hat{V}_{33} & \ddots
    & \\
    \vdots & \vdots &  & \ddots & \ddots & \\
    \hline
    &  & 0 &  &  & \begin{array}{c}
      \tmop{Hamiltonian}\\
      \tmop{elements} \tmop{between}\\
      \tmop{states} | \alpha \rangle
    \end{array}
  \end{array}\right) \label{H}
\end{equation}
where $\hat{V}_{n, m}$ are projections of $\hat{V}$ between states $| n^i
\rangle$ and $| m^j \rangle$ and $\hat{E}_n$ are projections of $\hat{H}_0$
between states $| n^i \rangle$. We can now use the standard degenerate
perturbation theory to show which Hamiltonian elements contribute to the
$m$-th order correction of the desired subspace. For small perturbation
$\lambda$ Hamiltonian can be written as:
\begin{equation}
  \hat{H} = \hat{H}_0 + \lambda \sum_n \mathbbm{P}_n \hat{V} \mathbbm{P}_n +
  \lambda^2 \sum_n \sum_{k \neq n} \frac{\mathbbm{P}_n \hat{V} \mathbbm{P}_k
  \hat{V} \mathbbm{P}_n}{E_n - E_k} + \cdots
\end{equation}
In general, $m$-th order correction will include matrices of the form:
\nolinebreak
\begin{equation}
  \mathbbm{P}_n \hat{V} \mathbbm{P}_{k_1} \hat{V} \mathbbm{P}_{k_2} \cdots
  \hat{V} \mathbbm{P}_{k_{m - 1}} \hat{V} \mathbbm{P}_n
\end{equation}
Looking at equation (\ref{H}), we can see that:
\begin{equation}
  \mathbbm{P}_n \hat{V} \mathbbm{P}_m = \left\{ \begin{array}{ll}
    \hat{V}_{n, n} & \tmop{if} m = n\\
    \hat{V}_{n, n + 1} & \tmop{if} m = n + 1\\
    \hat{V}_{n - 1, n}^T & \tmop{if} m = n - 1\\
    0 & \tmop{otherwise}
  \end{array} \right.
\end{equation}
Thus we can immediately conclude that for perturbation correction of order $m$
we need the following matrices:
\begin{equation}
  \underbrace{\hat{E}_0 + \lambda \hat{V}_{00} \nocomma, \hspace{1em} \lambda
  \hat{V}_{01}, \hspace{1em} \hat{E}_1 + \lambda \hat{V}_{11}, \hspace{1em}
  \ldots, \hspace{1em} \hat{E}_p + \lambda \hat{V}_{p p}, \hspace{1em}
  (\lambda \hat{V}_{p, p + 1})}_{m \tmop{matrices}}
\end{equation}
This means that in every step of Hamer's procedure, by including more states
in the Hamiltonian matrix, we increase the accuracy of the desired subspace of
states by two perturbation orders. More strictly, in SCE step $k$ we will have
precision of ground state energies up to order $2 k + 1$.

\section{Results and comparison}

The method described above was used on the generalized $t$-$V$ model with
various Mott insulating densities (critical densities). For a Mott insulator
the subspace of unperturbed ground states is very small
{\cite{Gomez-Santos1993}} and the Hamiltonian can be diagonalized
analytically.

\subsection{$Q = 1 / 2$ (half-filling), $p = 1$ (integrable), SCE step 3}

The truncated Hamiltonian for this case is of dimension $16 \times 16$, but for a very large system size $L$ it
can be separated into two equal subspaces of dimension $8 \times 8$, which can
be easily diagonalized. The condition for the system size is $L > (2 \times \text{step} + 1)(p+1)$.
The ground state is therefore 2-fold degenerate and
the ground state energy was calculated to be:
\begin{equation}
  E_0 = - \frac{L}{U}t^2 + \frac{L }{U^3}t^4 + \mathcal{O} (t^8)
\end{equation}
The density-density correlation functions $N_m = \left\langle \sum_{i = 1}^L
\hat{n}_i \hat{n}_{i + m} \right\rangle$ were found to be:
\footnotesize
\begin{multicols}{2}
\begin{equation}
  N_1 = L \frac{t^2}{U^2} - 3 L \frac{t^4}{U^4} + \mathcal{O}(t^8)
\end{equation}
\begin{equation}
  N_2 = \frac{L}{2} - 2 L \frac{t^2}{U^2} + 7 L \frac{t^4}{U^4} + \mathcal{O}(t^6)
\end{equation}
\begin{equation}
  N_3 = 2 L \frac{t^2}{U^2} - 5 L \frac{t^4}{U^4} + \mathcal{O} (t^6)
\end{equation}
\begin{equation}
  N_4 = \frac{L}{2} - 2 L \frac{t^2}{U^2} + 2 L \frac{t^4}{U^4} + \mathcal{O} (t^6)
\end{equation}
\begin{equation}
  N_5 = 2 L \frac{t^2}{U^2} - 2 L \frac{t^4}{U^4} + \mathcal{O} (t^6)
\end{equation}
\end{multicols}
\normalsize
This particular case of the generalized $t$-$V$ model can be mapped to the
Heisenberg XXZ spin model with background magnetic field, which is solved
analytically by Orbach {\cite{Orbach1958}} and Walker {\cite{Walker1959}}. On
closer inspection we can see that the analytical expansions of ground state
energy and density-density correlator $N_1$ (in the language of spins this is the
spin-spin correlator) presented in {\cite{Walker1959}} match our results.

Furthermore, the XXZ model for $\frac{t}{U} \rightarrow 0$ is equivalent to
the Ising model {\cite{Takahashi2005}} for which the long-range
density-density correlators are:
\begin{equation}
  N_m = \left\{ \begin{array}{ll}
    0 & \tmop{for} m \tmop{odd}\\
    \frac{L}{2} & \tmop{for} m \tmop{even}
  \end{array} \right. \label{isingdensity}
\end{equation}
which is fully consistent with our results.

The current density is given by:
\begin{equation}
	J = - i t \left\langle \sum_{i = 1}^L
	\hat{c}_i^\dagger \hat{c}_{i + 1} - \text{h.c.}\right\rangle
\end{equation}
and was found to be zero up to order $\mathcal{O} (t^8)$ for large systems.

Model for $p = 1$ was inspected thoroughly in the first order approximation in
Refs. {\cite{Gomez-Santos1993,Dias2000}} where the ground state energy and the
current density should both vanish for the half-filling case, which also
agrees with our results.

\subsection{$Q = 1 / 3$, $p = 2$ (non-integrable), SCE step 3}

For $p > 1$ the model is non-integrable. In step 3 (7th order of
perturbation), the Hamiltonian is of dimension $36 \times 36$, however it can
be divided into three equivalent subspaces of dimension $12 \times 12$. The
ground state is therefore 3-fold degenerate and its energy was found to be:
\begin{equation}
  E_0 = - \frac{2 L}{3 U_2} t^2 + \left( \frac{2 L}{3 U_2^3} - \frac{2 L}{U_1
  U_2^2} \right) t^4 + \left( \frac{16 L}{3 U_1 U_2^4} - \frac{17 L}{3 U_1^2
  U_2^3} - \frac{10 L}{3 U_1^3 U_2^2} \right) t^6 + \mathcal{O} (t^8)
\end{equation}
The density-density correlators are:
\footnotesize
\begin{equation}
  N_1 = \frac{2 L}{U_1^2 U_2^2} t^4 + \left( \frac{10 L}{U_1^4 U_2^2} +
  \frac{34 L}{3 U_1^3 U_2^3} - \frac{16 L}{3 U_1^2 U_2^4} \right) t^6 + \mathcal{O}
  (t^8)
\end{equation}
\begin{equation}
  N_2 = \frac{2 L}{3 U_2^2} t^2 + \left( \frac{4 L}{U_1 U_2^3} - \frac{2
  L}{U_2^4} \right) t^4 + \left( \frac{20 L}{3 U_1^3 U_2^3} + \frac{17
  L}{U_1^2 U_2^4} - \frac{64 L}{3 U_1 U_2^5} \right) t^6 + \mathcal{O} (t^8)
\end{equation}
\begin{equation}
  N_3 = \frac{L}{3} - \frac{4 L}{3 U_2^2} t^2 + \left( - \frac{16 L}{3 U_1^2
  U_2^2} - \frac{8 L}{U_1 U_2^3} + \frac{13 L}{3 U_2^4} \right) t^4 + \mathcal{O} (t^6)
\end{equation}
\begin{equation}
  N_4 = \frac{2 L}{3 U_2^2} t^2 + \left( \frac{10 L}{3 U_1^2 U_2^2} + \frac{4
  L}{U_1 U_2^3} - \frac{7 L}{3 U_2^4} \right) t^4 + \mathcal{O} (t^6)
\end{equation}
\begin{equation}
  N_5 = \frac{2 L}{3 U_2^2} t^2 + \left( \frac{10 L}{3 U_1^2 U_2^2} + \frac{4
  L}{U_1 U_2^3} - \frac{L}{3 U_2^4} \right) t^4 + \mathcal{O} (t^6)
\end{equation}
\normalsize
Similarly to equation (\ref{isingdensity}), we expect that for $Q = \frac{1}{p
+ 1}$ the density-density correlation functions in the limit of $\frac{t}{U_m}
\rightarrow 0$ to be:
\begin{equation}
  N_m = \left\{ \begin{array}{ll}
    \frac{L}{p} & \tmop{for} m \tmop{divisible} \tmop{by} p\\
    0 & \tmop{otherwise}
  \end{array} \right. \label{correlationsgeneral}
\end{equation}
and it is indeed true for our results.

Again, the current density is zero up to order $\mathcal{O} (t^8)$ for large systems.

\subsection{$Q = 1 / 4$, $p = 3$ (non-integrable), SCE step 3}

This is another non-integrable case. The Hamiltonian is of dimension $52
\times 52$, but it consists of four equal subspaces of dimension $13 \times
13$. The ground state is thus 4-fold degenerate and has energy:
\footnotesize
\begin{equation}
  E_0 = - \frac{L}{2 U_3} t^2 + \left( \frac{L}{2 U_3^3} - \frac{3 L}{2 U_2
  U_3^2} \right) t^4 + \left( \frac{4 L}{U_2 U_3^4} - \frac{17 L}{4 U_2^2
  U_3^3} - \frac{5 L}{2 U_2^3 U_3^2} - \frac{5 L}{U_1 U_2^2 U_3^2} \right) t^6
  + \mathcal{O} (t^8)
\end{equation}
\normalsize
The density-density correlation functions are:
\footnotesize
\begin{equation}
  N_1 = \frac{5 L}{U_1^2 U_2^2 U_3^2} t^6 + \mathcal{O} (t^8)
\end{equation}
\begin{equation}
  N_2 = \frac{3 L}{2 U_2^2 U_3^2} t^4 + L \left( \frac{15}{2 U_2^4 U_3^2} +
  \frac{17}{2 U_2^3 U_3^3} - \frac{4}{U_2^2 U_3^4} + \frac{10}{U_1 U_2^3
  U_3^2} \right) t^6 + \mathcal{O} (t^8)
\end{equation}
\begin{equation}
  N_3 = \frac{L}{2 U_3^2} t^2 - L \left( \frac{3}{2 U_3^4} - \frac{3}{U_2
  U_3^3} \right) t^4 + L \left( \frac{5}{U_2^3 U_3^3} + \frac{51}{4 U_2^2
  U_3^4} - \frac{16}{U_2 U_3^5} + \frac{10}{U_1 U_2^2 U_3^3} \right) t^6 + \mathcal{O}
  (t^8)
\end{equation}
\begin{equation}
  N_4 = \frac{L}{4} - \frac{L}{U_3^2} t^2 + L \left( \frac{13}{4 U_3^4} -
  \frac{4}{U_2^2 U_3^2} - \frac{6}{U_2 U_3^3} \right) t^4 + \mathcal{O} (t^6)
\end{equation}
\begin{equation}
  N_5 = \frac{L}{2 U_3^2} t^2 + L \left( \frac{2}{U_2^2 U_3^2} + \frac{3}{U_2
  U_3^3} - \frac{2}{U_3^4} \right) t^4 + \mathcal{O} (t^6)
\end{equation}
\normalsize
Again, our results for correlators are consistent with equation
(\ref{correlationsgeneral}).

For a large system size the current density was calculated to be zero up to
perturbation order $\mathcal{O} (t^8)$.

\section{Summary \& outlook}

We have shown that the strong coupling expansion devised for numerically
solving lattice quantum field theory problems can also be used in the field of
quantum spin models, giving us analytical results. Our test model was the
long-range $t$-$V$ model at critical densities. For the integrable system (XXZ
model) our results are fully consistent with previous work and for the
non-integrable models we have obtained various observables not obtained
before.

The next step will be to expand this method to near-critical densities, where
there is one additional hole in the system, two additional holes, etc. Though
the initial subspace will probably be too big to use analytics, it should be
small enough to use numerical approach. The preliminary results show that for
a system $p = 2$ (non-integrable), $L = 3 N + 1$ (one hole), the Hamiltonian has dimension $2 L
\times 2 L$ in the SCE step 2 and the ground state energy is:
\begin{equation}
  E_0 = \left\{ \begin{array}{ll}
    - 2 t - \frac{2 N}{U_2} t^2 + \frac{2}{U_2^2} t^3 + \mathcal{O} ( t^4 ) &
    \text{for odd } N\\
    - 2 t \cos \frac{\pi}{L} - A_{(L)} \frac{2 N}{U_2} t^2
    + B_{(L)} \frac{2}{U_2^2} t^3 + \mathcal{O} ( t^4)
		& \text{for even } N
  \end{array} \right.
\end{equation}
with functions $A_{(L)}$ and $B_{(L)}$ that can be numerically approximated as
$A_{(L)} = 1 - \frac{64.1 (4)}{L^3}$ and $B_{(L)} = 1 + \frac{68 (1)}{L^2} + \frac{1\underline{2}00 (1)}{L^4}$. Further investigation is needed.

The assumption $U_m \leqslant \frac{U_{m - 1} + U_{m + 1}}{2}$ that was introduced
in the beginning could be also abandoned. The system will now have different
phases, depending on the potential energies $U_m$. However, SCE approach should still work if
the initial subspace of states is properly chosen for the specific system setup.

\ack
M.S. is fully funded by EPSRC, NoWNano DTC grant number EP/G03737X/1.
E.B. acknowledges partial support by the ERC grant 279738-NEDFOQ and by Lancaster University via
ECSG grant SGS/18/01.

\end{document}